\documentclass{article}

\usepackage[accepted]{icml2019}
\usepackage{amsmath}
\usepackage{amssymb}
\usepackage{nccmath}
\usepackage{mathtools}
\usepackage{graphicx} 
\usepackage{hyperref}

\usepackage{amsthm}
\usepackage{enumitem}
\usepackage{mwe}
\usepackage{subfig}

\begin{document}

\twocolumn[
\icmltitle{Directivity Modes of Earthquake Populations with Unsupervised Learning}

\begin{icmlauthorlist}
\icmlauthor{Zachary E. Ross}{Caltech_SL}
\icmlauthor{Daniel T. Trugman}{LANL}
\icmlauthor{Kamyar Azizzadenesheli}{Caltech_CMS}
\icmlauthor{Anima Anandkumar}{Caltech_CMS}{}
\end{icmlauthorlist}

\icmlaffiliation{Caltech_SL}{Seismological Laboratory, California Institute of Technology, Pasadena, CA 91125}
\icmlaffiliation{Caltech_CMS}{Dept. of Computing and Mathematical Sciences, California Institute of Technology, Pasadena, CA 91125}
\icmlaffiliation{LANL}{Geophysics Group, Earth and Environmental Sciences Division, Los Alamos National Laboratory, Los Alamos, NM 87545}
\icmlcorrespondingauthor{Zachary E. Ross}{zross@gps.caltech.edu}


\DeclarePairedDelimiter\abs{\lvert}{\rvert}%
\DeclarePairedDelimiter\norm{\lVert}{\rVert}%
\makeatletter
\let\oldabs\abs
\def\abs{\@ifstar{\oldabs}{\oldabs*}}
\let\oldnorm\norm
\def\norm{\@ifstar{\oldnorm}{\oldnorm*}}
\makeatother

\vskip 0.3in
]
\printAffiliationsAndNotice{}

\begin{abstract}
We present a novel approach for resolving modes of rupture directivity in large populations of earthquakes. A seismic spectral decomposition technique is used to first produce relative measurements of radiated energy for earthquakes in a spatially-compact cluster. The azimuthal distribution of energy for each earthquake is then assumed to result from one of several distinct modes of rupture propagation. Rather than fitting a kinematic rupture model to determine the most likely mode of rupture propagation, we instead treat the modes as latent variables and learn them with a Gaussian mixture model. The mixture model simultaneously determines the number of events that best identify with each mode. The technique is demonstrated on four datasets in California with several thousand earthquakes. We show that the datasets naturally decompose into distinct rupture propagation modes that correspond to different rupture directions, and the fault plane is unambiguously identified for all cases. We find that these small earthquakes exhibit unilateral ruptures 53-74\% of the time on average. The results provide important observational constraints on the physics of earthquakes and faults.

\end{abstract}

\section{Introduction}

\begin{figure*}[t]
\centering
\includegraphics[width=\textwidth]{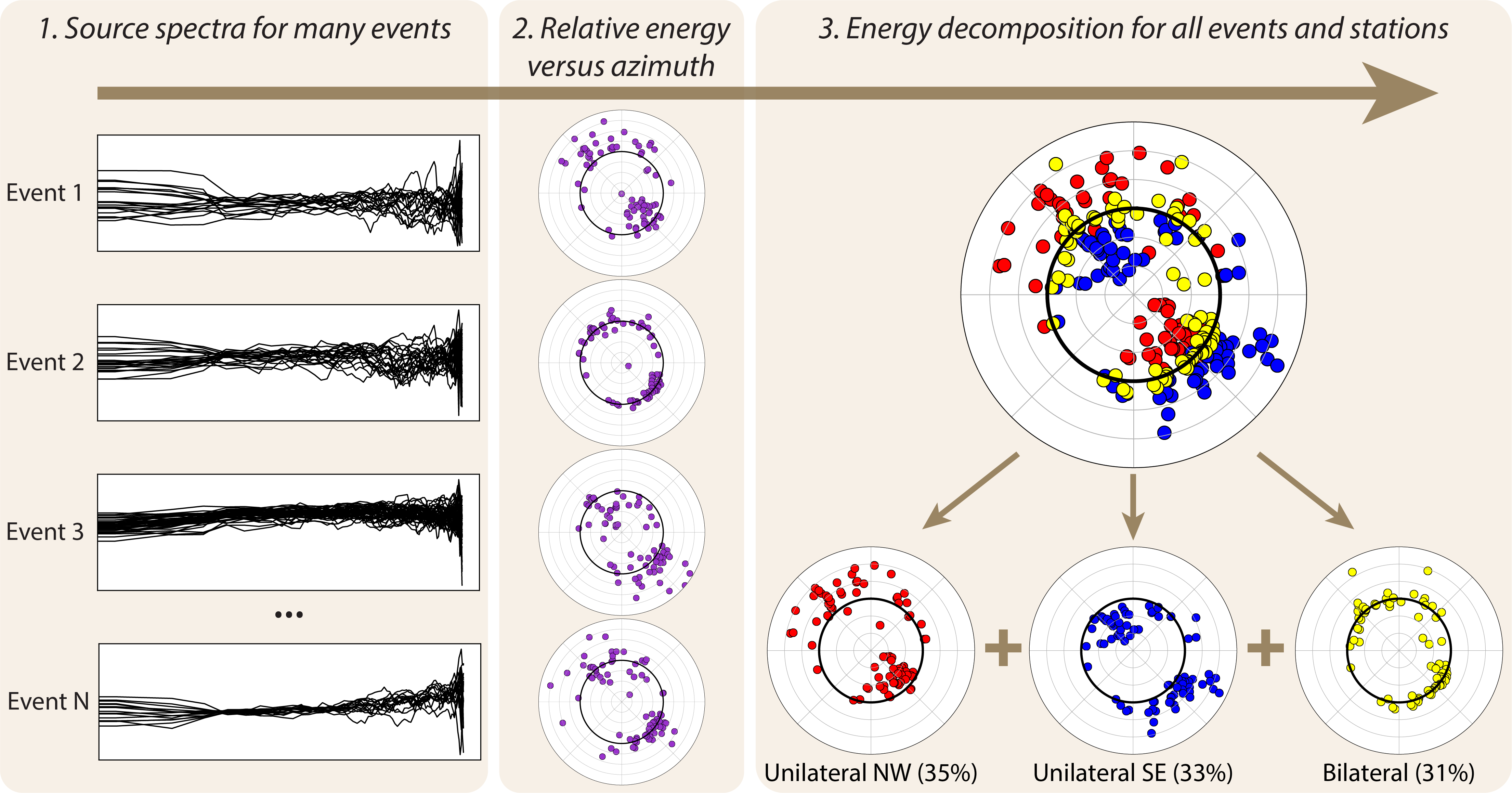}
\caption{Earthquake source spectra are calculated for many events and stations in a compact cluster. Then, relative energy values are calculated for all stations and events. The set of energy values for all earthquakes is decomposed into $K$ distinct rupture propagation modes. The decomposition simultaneously determines the fraction of events that belong to each mode.}
\label{fig:example}
\end{figure*}

The kinematics of the earthquake source process has a first-order effect on seismic ground motions. In particular, it is well-known that for moving sources, rupture directivity induces azimuthally dependent changes in the frequency content of seismic waves \citep{haskell1964total}. Often earthquake ruptures are simplified as falling into one of two end-member modes:  bilateral (symmetric) ruptures and fully unilateral ruptures. These rupture modes produce distinct signatures in the seismic radiation. Some studies discuss a continuum of scenarios in between these modes for which there is some degree of asymmetry in the distribution of seismic moment \citep[e.g.][]{boatwright1984effect}.

Theoretical studies of earthquake physics have suggested various links between rupture propagation and properties of fault zones. For example, large earthquakes that saturate the seismogenic zone are often viewed as being constrained geometrically to have elongated ruptures \citep{mcguire2002predominance}. The bimaterial hypothesis \citep{weertman1980unstable, andrews1997wrinkle} predicts that for faults with a velocity contrast across them, a rupture direction that is aligned with the direction of slip in the more compliant medium will be favored due to a dynamic reduction of normal stress at the crack tip. Other models may anticipate earthquakes to rupture with unilateral directivity as a frictional phenomenon, such as when the nucleation size is much smaller than the size of a seismogenic patch \citep[e.g.][]{michel2017pulse,lin2018microseismicity}. It is desirable to better understand whether any of these models are relevant to earthquakes in nature, and whether there are resolvable differences in the properties of faults that can lead to different physics. Providing observational constraints on some of these physical models would require assembling directivity measurements for large numbers of earthquakes in order to perform a statistical analysis.

The extent of rupture directivity is readily determined for M $\geq 6$ earthquakes with various seismological methods \citep[e.g.][]{mcguire2001teleseismic, ye2016rupture, van2018improved}. These events are recorded typically by many stations and are sufficiently large that a slip model can be obtained using seismic waveforms recorded at teleseismic distances. However, the number of these large events on any single fault is too few to determine the statistical properties of rupture directivity patterns. Small earthquakes on the other hand are plentiful on individual faults due to the well-known power law magnitude distribution. There have been various approaches to estimating rupture directivity in small earthquakes, such as the use of higher-order moment tensors \citep{mcguire2004estimating}, fitting directivity functions \citep{boatwright2007persistence, tan2010rupture, wang2011rupture, kane2013rupture, abercrombie2017earthquake}, measuring spectral splitting \citep{ross2016toward, calderoni2015along}, measuring variations in apparent duration among repeating earthquake sequences \citep{lengline2011rupture}, and azimuthal analyses of ground motion \citep{kurzon2014ground}. For most of these approaches, empirical Green's functions are needed to correct the data for propagation and site effects \citep[e.g.][]{mueller1985source, hough1995source, prieto2004earthquake}, and then some kind of geophysical inverse problem is solved to obtain measurements about directivity.

In this paper, we present a data-driven approach to learning modes of directivity in earthquake populations that can scale to large datasets (Fig. \ref{fig:example}). We demonstrate that for a population of earthquakes, the azimuthal distribution of radiated energy naturally decomposes into distinct modes of rupture propagation without assuming a fault plane geometry or solving a geophysical inverse problem. The modes generally represent unilateral ruptures in different directions as well as bilateral ruptures. This decomposition simultaneously determines the fraction of events in each mode. We demonstrate the method on four datasets from California with thousands of events each, and clearly identify the fault plane and rupture directivity modes for all four cases. We show that the earthquakes in these clusters have predominantly unilateral ruptures, with bilateral ruptures being relatively infrequent.

\section{Methods}

Our approach to resolving directivity modes is based on the idea that an individual earthquake within a population can be represented as a random sample from one of a handful of end-member rupture scenarios. For strike-slip faults, the number of observable cases might be three: a bilateral rupture with symmetric rupture propagation, and two unilateral rupture modes that propagate in opposite directions along a fault. Alternatively, one might want to only consider unilateral directivity modes, with bilateral ruptures corresponding to a superposition of unilateral modes. The problem can then be formulated as one of recovering the distinct rupture modes that exist in the data. Our approach can be summarized with the following steps:

\begin{enumerate}
    \item A cluster of earthquakes is identified for detailed analysis
    \item The seismograms from these earthquakes are used with a seismic spectral decomposition algorithm to obtain apparent source spectra for as many events and stations as possible (Fig. \ref{fig:example}, step 1).
    \item The radiated energy at each station is determined by integrating the square of the velocity spectrum (Fig. \ref{fig:example}, step 2).
    \item We apply an algorithm for filling in missing data values by learning a low-rank approximation to the entire dataset.
    \item Directivity modes are learned with a Gaussian mixture model. (Fig. \ref{fig:example}, step 3)
    \item Each mode can be visually interpreted as an azimuthal directivity function for a rupture mode that exists repeatedly in the data. The modes are obtained without solving a geophysical inverse problem.
\end{enumerate}

We now discuss each of these steps in detail. 

\subsection{Data Pre-processing and Quality Control}
The first step of the procedure involves selecting a cluster of earthquakes for analysis of rupture propagation. In this study, we use spatially-compact clusters for which the source-receiver azimuths are roughly the same for all events. We restrict the minimum source-receiver distance to 20 km to help ensure this approximation is valid. 
For each event, we obtain all available automated and manual P-wave picks, and for each associated waveform, we use these picks to define a 1.5~s signal window starting 0.1s before the listed P-wave arrival, and a background noise window of equal length immediately preceding it. We use a multitaper algorithm \citep{prieto_fortran_2009,krischer_mtspec_2016} to compute seismic amplitude spectra on all available vertical-component channels with sampling rates $\geq$100~Hz. We convert each spectrum to units of displacement and interpolate where necessary to a uniform frequency spacing of 2/3~Hz with minimum and maximum values of 0~Hz and 50~Hz (the Nyquist frequency for 100~Hz sampling). 

For the purposes of quality control, we further consider only spectra with signal-to-noise ratio (SNR) greater than 5 within three frequency bands spanning the 3~Hz to 30~Hz range. Clipped waveforms are commonly observed on short-period and broadband stations for moderate earthquakes, and to mitigate this we further exclude spectra flagged by an automated clipped detection algorithm based on the fourth moment of the time domain waveform amplitudes observed in the signal window \citep{trugman_application_2017}. The remaining displacement spectra after these waveform pre-processing steps and quality control steps are the inputs for the spectral decomposition method described below.

\subsection{Spectral Decomposition}

Spectral decomposition is a technique designed to separate the observed displacement spectra $d(f)$ of earthquakes $i$ recorded and a set of seismic stations $j$ into source, path, and site terms. As described in detail by \citet{trugman_application_2017}, for densely-recorded datasets, each earthquake will be recorded by many stations, each approximate source-station path will be sampled many times, and each station will record many earthquakes. If these conditions hold, then by working in the log frequency domain, relative source terms $s_i$, path terms $p_{k(i,j)}$, and site terms $st_j$ can be estimated at each frequency point as part of an overdetermined inverse problem defined by the linear equation:
\begin{equation}
\label{eq:decomposition}
    d_{ij}(f) = s_i(f) + p_{ij}(f) + st_j(f) + \epsilon_{ij}(f).
\end{equation}
Following \citet{shearer_comprehensive_2006} and \citet{trugman_application_2017}, we assume azimuthally isotropic path terms $p$ that depend only on the source-receiver travel-time. We limit our analysis to stations with epicentral distances less than 100~km, and to ensure that the basic assumptions of spectral decomposition hold, we require each station used in the analysis to have recorded at least 20 earthquakes. We estimate the source, path, and site terms defined by equation (\ref{eq:decomposition}) using an iterative, robust least-squares inversion that uses Huber-norm weighting to suppress the influence of outliers on the final solution \citep{shearer_comprehensive_2006}.

Our primary observational constraints come not from the source spectra $s(f)$ themselves, but instead the apparent source spectra $sa(f)$ observed at each station and how they vary as a function of azimuth:
\begin{equation}
    sa_{ij}(f) = d_{ij}(f) - p_{ij}(f) - st_j(f).
\end{equation}
The apparent source spectrum $sa(f)$ is thus a path- and site-corrected form of the observed displacement spectrum, while the source spectrum $s(f)$ itself is the azimuthal average of apparent source spectrum across all stations (to good approximation). Rupture directivity can cause significant azimuthal variations in apparent spectra, with higher amplitudes along azimuths aligned with rupture direction. One general limitation of the spectral decomposition method is that the estimated terms only resolve relative differences between each source, each path, and each site, as one can add an arbitrary function to all source terms and subtract the same function from all site or path terms without effecting the misfit. While this presents a challenge for estimating earthquake source parameters like corner frequency and stress drop \citep{shearer_comparing_2019}, it does not affect the results presented in this study, which are based upon on the relative variations in spectral energy as a function of azimuth, and not the absolute energy values.

\subsection{Building the Data Matrix}
By this point, we have obtained apparent source spectra for many events at many stations. From here, we proceed to calculate apparent radiated energy values for the spectra. For an apparent source displacement spectrum, $sa(f)$, the apparent radiated energy (up a multiplicative constant) can be calculated as,

\begin{equation}
    E_R = 4 \pi^2  \int_{f_{min}}^{f_{max}} f^2 \lvert sa(f)\lvert^2\, df.
\end{equation}

The optimal values of $f_{min}$ and $f_{max}$ generally depend on the instrument response, sampling rate, and the magnitude range of the events for which the directivity analysis will be performed. They further depend on the signal to noise ratio of the spectra. Our spectra generally have low SNR below 3 Hz and above 30 Hz, so we set $f_{min}$ and $f_{max}$ equal to these values.

Next, we normalize the $E_R$ values separately for each event by dividing by the median value calculated over all of the stations. This results in a set of values that indicate whether $E_R$ at a given station is greater than or less the median. In doing so, we require a minimum of 10 stations for which $E_R$ values are available, or else the event is skipped altogether. We then remove all stations from the dataset which have fewer than 20 spectra over all the events. Then, we identify outlier values by calculating the median absolute deviation (in log units) and looking to see whether there are any values for a single event that are more than 5 deviations away from the median. If so, these values are removed from the dataset. If the outlier removal process results in fewer than 10 values for a single event, the event is skipped.

Applying the steps described in this section results in a set of relative $E_R$ values for many events and stations. However, of the complete set of stations available, few if any events will have $E_R$ values at all of them. This is especially true for the smallest events in the dataset, which are much more frequent than the larger ones. However, to fit a latent variable model to the data, there cannot be any missing values, and therefore they must be filled.

\begin{figure}[tb]
\centering
\includegraphics[width=\columnwidth]{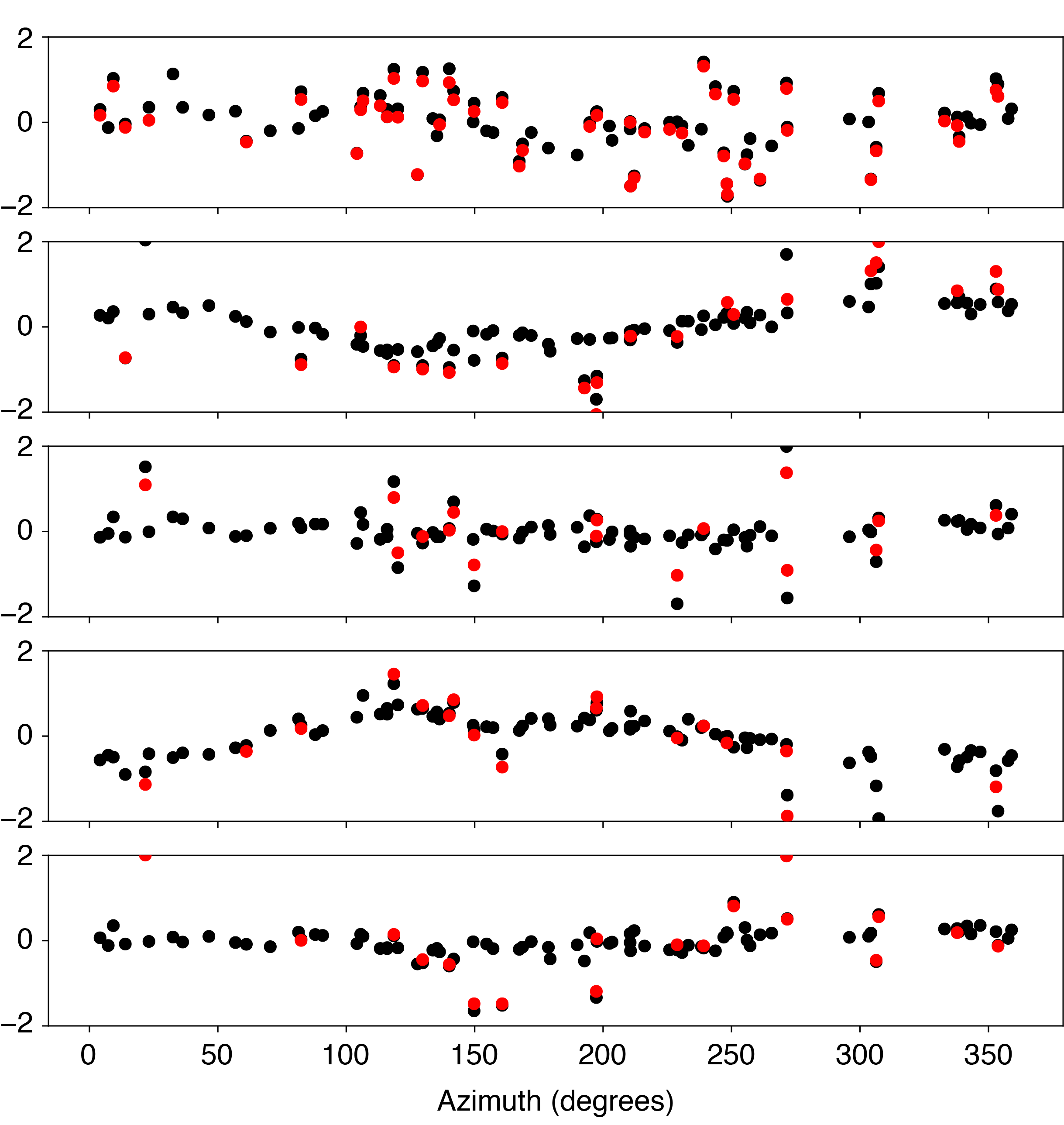}
\caption{Examples of relative $\log{E_R}$ distributions for five earthquakes. The data are shown before (red circles) and after (black circles) applying the imputation algorithm.}
\label{fig:imputation_ex}
\end{figure}

In applied mathematics there is an extensive literature on data imputation algorithms for filling missing data. The simplest form of these algorithms fills missing values with the mean or median for each variable. Other algorithms are more sophisticated, trying to learn a lower dimensional representation of the data from the observations that can be used to reconstruct the missing values. Algorithms of this type have seen some use in geophysics, for example to fill gaps in GPS data prior to performing principal component analysis or independent component analysis \citep[e.g.][]{kositsky2010inverting}. In this study, we chose an imputation algorithm that learns a low-rank approximation to the existing data, which can then be used to fill in missing values. This technique is based on thresholding a singular value decomposition of the data \citep{hastie_matrix_2014}. Some examples of applying this technique are shown in Figure \ref{fig:imputation_ex}, where real values are shown alongside imputed values. Filled values have a tendency to be generally close to zero (the mean) unless there is significant evidence in the data otherwise.

\subsection{Gaussian mixture models}
Gaussian mixture models (GMM) are a type of latent variable model that assumes a generative process can be represented as a superposition of two or more Gaussian distributions. Each Gaussian represents a distinct mode of occurrence, and there is a probability $w_i$ that a realization $x \in \mathbb{R}^d$ will be drawn from mode $i$. Specifically, the model can be written as,

\begin{equation}
    p(x) = \sum_{k=1}^K w_k \, \mathcal{N}(x | \mu_k, \Sigma_k)
\end{equation}
\begin{equation}
    \sum_{k=1}^K w_k = 1
\end{equation}
where $K$ is the number of components in the GMM and $\mathcal{N}(x | \mu_k, \Sigma_k)$ is the multivariate Gaussian distribution. Since $\mu_k$ and $\Sigma_k$ are not directly observed, they are latent variables; however they can be recovered by fitting a GMM to some data. The most common approach to estimating the parameters of a GMM is with the Expectation-Maximization algorithm \citep{dempster1977maximum}, which is the method we use in this paper. We also tested the method of moments algorithm \citep{anandkumar2014tensor} that matches the empirical moments to the theoretical ones through a tensor decomposition algorithm. We found that the EM algorithm was more effective at separating the modes for the datasets considered in this study.

Our measured $\log{E_R}$ values for each earthquake correspond to the $x$ in the GMM, and we therefore wish to obtain the $\mu_k$ and $\Sigma_k$ for the $K$ modes in the data.  For this paper, we will work with strike-slip faults and consider both $K=2$ and $K=3$ cases. The $K=2$ scenario would ideally recover two unilateral directivity modes, while the $K=3$ scenario would correspond to a bilateral rupture mode and two opposite unilateral modes.  We will further assume that the covariances are spherical, i.e. $\Sigma_k=\sigma_k^2I$. With this formulation, each $\mu_{k}$ is a vector with dimensionality equal to the number of stations used in the analysis, and the measured $\log{E_R}$ of an individual earthquake are modeled as a random sample from the GMM.

\subsection{Uncertainty Analysis}

To estimate the uncertainty in the model parameters, we use a bootstrapping approach. We resample the events with replacement 1000 times and re-fit the model each time. Since the order of the elements of $w$ is randomly determined when the model is fit, we obtain all permutations of the columns of $\mu$ and calculate the $\ell_1$ norm of the difference between the best fitting $\mu^\text{best}$ and each of the permutations. The permutation with the smallest $\ell_1$ value is taken as the $w$ for that particular resampling. Then we calculate a 95\% confidence interval (CI) for each of the $w_k$.

\section{Related Work}
In this study, our goal is to estimate the fraction of events for each rupture directivity mode. We are interested in working with hundreds to thousands of earthquakes at a time to ensure that the statistics are robust. To date, there are several studies that have examined directivity on such a scale.

\citet{wang2011rupture} calculated spectral ratios using empirical Green's functions and fit three different models (bilateral and two unilateral) to each event separately to determine the most likely rupture mode. They perform this analysis in the creeping section of the San Andreas fault and test the method on more than 900 earthquakes. They concluded that roughly 40\% had bilateral ruptures, and of the unilateral ruptures, most had southeast directivity.

\citet{kane2013rupture} used a spectral decomposition technique to obtain apparent source spectra for thousands of events at the Parkfield section of the San Andreas fault. They fit a unilateral directivity model to the apparent spectra for each event individually, and concluded that there was slightly more events with southeast ruptures than northwest ruptures. They did not consider bilateral ruptures.

Building on the results of \citet{calderoni2015along} and \citet{pacor_diminishing_2016} for the L'Aquila, Italy earthquake sequence, \citet{calderoni_rupture_2017} used azimuthal variations in S-wave spectra to demonstrate that along-strike directivity is a common feature of normal faulting earthquakes in the central Appennines. Earthquakes from the Umbria Marche, L'Aquila, and 2016-2017 central Italy sequences exhibit temporally persistent and spatially coherent directivity patterns, which suggests that in-situ fault and geologic properties play an important role in the preferred rupture direction.

A different approach was developed by \citet{lengline2011rupture}, who exploited repeating earthquake sequences in Parkfield to simultaneously invert for relative perturbations to apparent duration for all earthquake pairs and stations. They used P-wave signals for their analysis, and their kinematic directivity model only allowed for unilateral ruptures. \citet{lengline2011rupture} concluded that the majority of earthquakes analyzed exhibited southeast rupture signals.

A time-domain method based on ground-motion prediction equations was developed by \citet{kurzon2014ground}. They proposed an empirical directivity index based on observed azimuthal variations in ground motion amplitude. They applied the method to two clusters of earthquakes in the central San Jacinto fault zone.

\section{Experiments}
Now, we demonstrate the method on four different datasets from California. These datasets use P-wave seismograms only and were chosen for several reasons. First, each dataset is a spatially-compact cluster with thousands of events. Second, two of them have been used in previous studies to analyze directivity and can serve as a point of reference.

\subsection{Cahuilla swarm}
\begin{figure}[tb]
\centering
\includegraphics[width=\columnwidth]{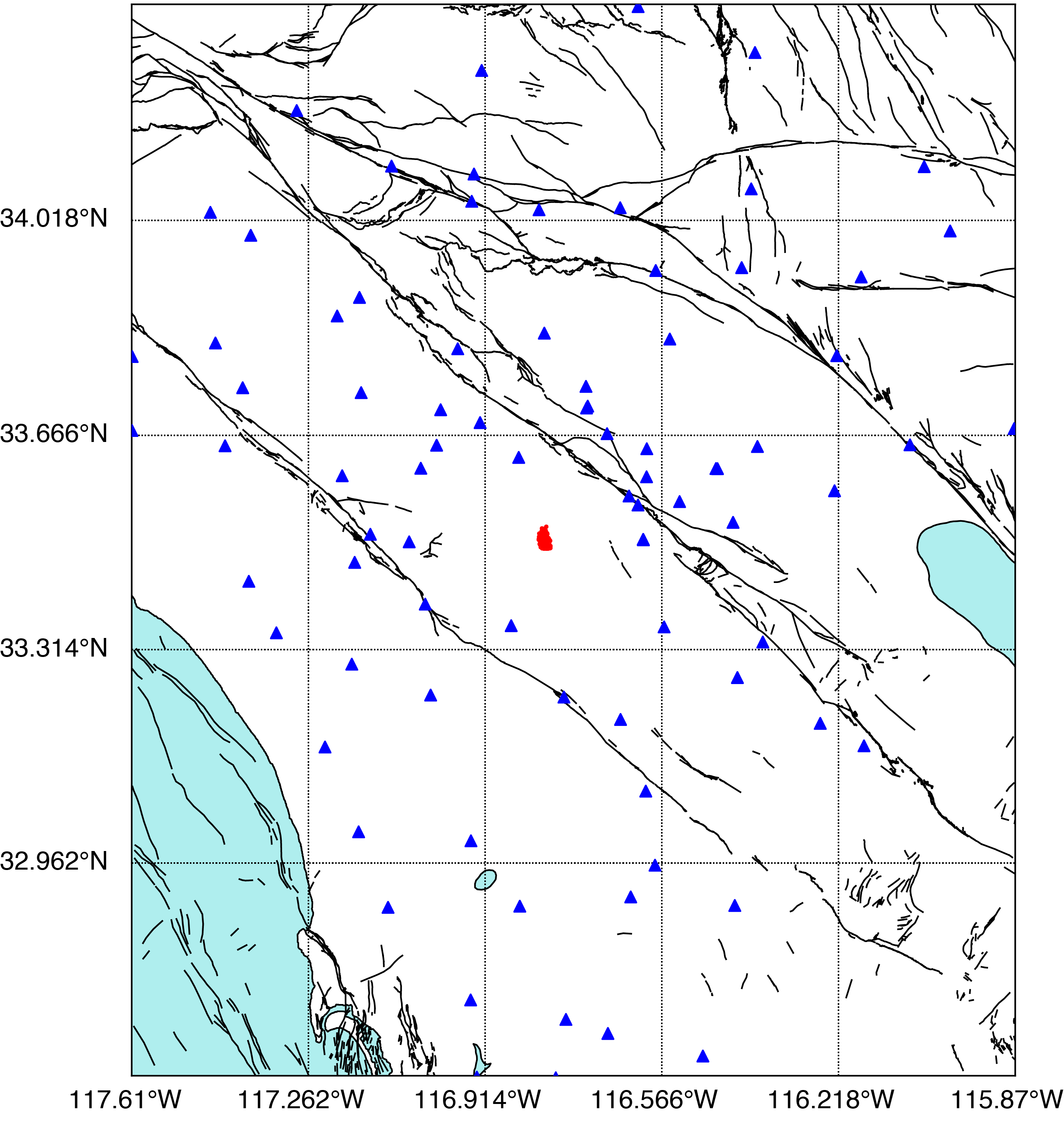}
\caption{Map of the Cahuilla swarm and surrounding region. Earthquakes are shown as red dots. Seismic stations used in this study are indicated by blue triangles.}
\label{fig:cahuilla_map}
\end{figure}

The first dataset contains seismograms for 11,631 events that occurred as part of the Cahuilla earthquake swarm in Southern California during 2016-2019 (\citet{hauksson2019slow}; Fig. \ref{fig:cahuilla_map}). The largest of these events is $M 4.4$, while most events are considerably smaller. These events were studied in detail by \citet{hauksson2019slow}, who observed that the events have right-lateral strike-slip focal mechanisms with generally little variation in strike. The waveform and meta data are all publicly available from the Southern California Earthquake Data Center. 

\begin{figure}[tb]
\centering
\includegraphics[width=\columnwidth]{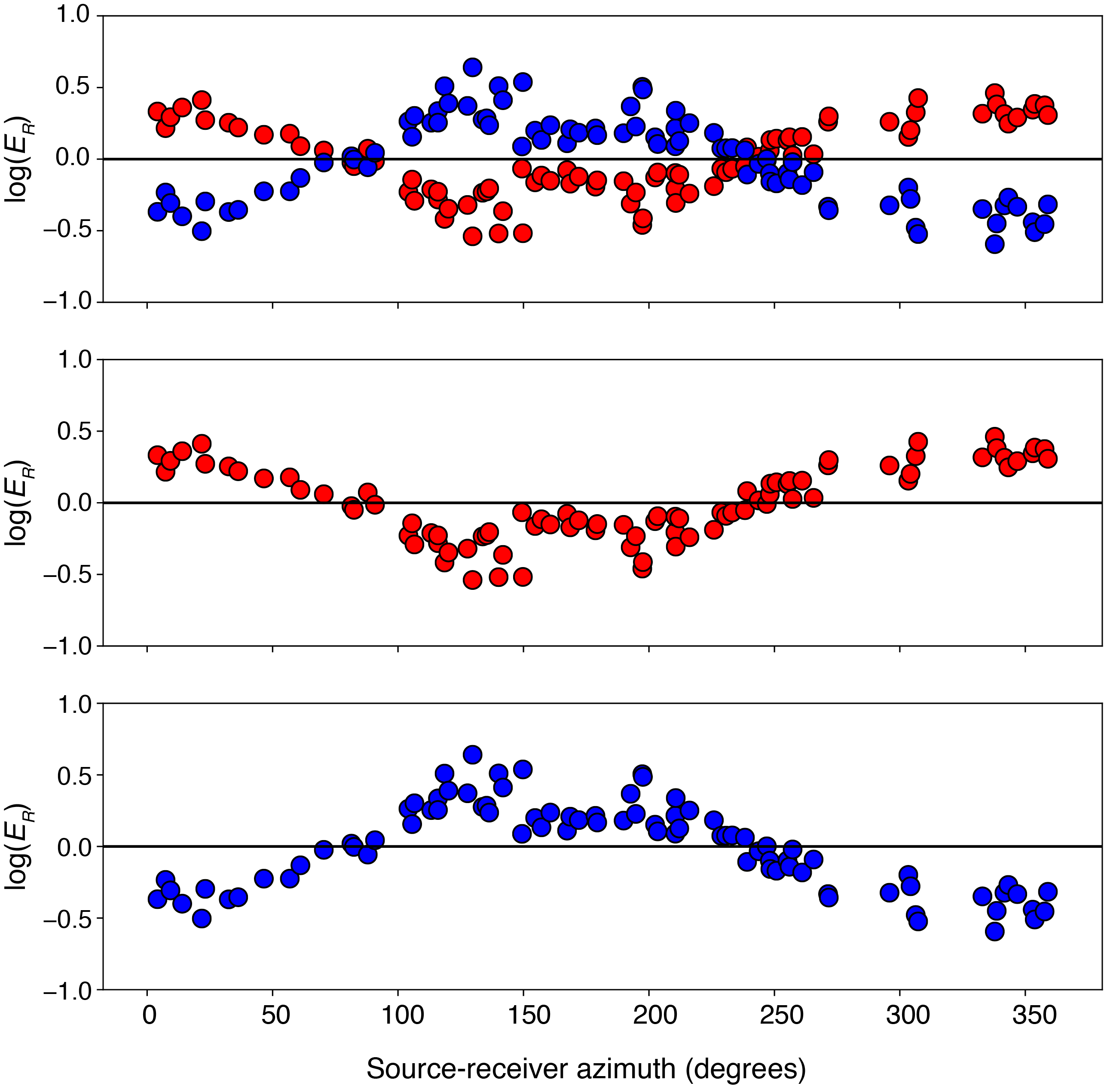}
\caption{Two-mode decomposition for the Cahuilla swarm. The modes correspond directly to the $w_k$ determined by fitting the GMM (Table \ref{table:cahuilla}).}
\label{fig:result_polar_cahuilla}
\end{figure}

\begin{figure}[tb]
\centering
\includegraphics[width=\columnwidth]{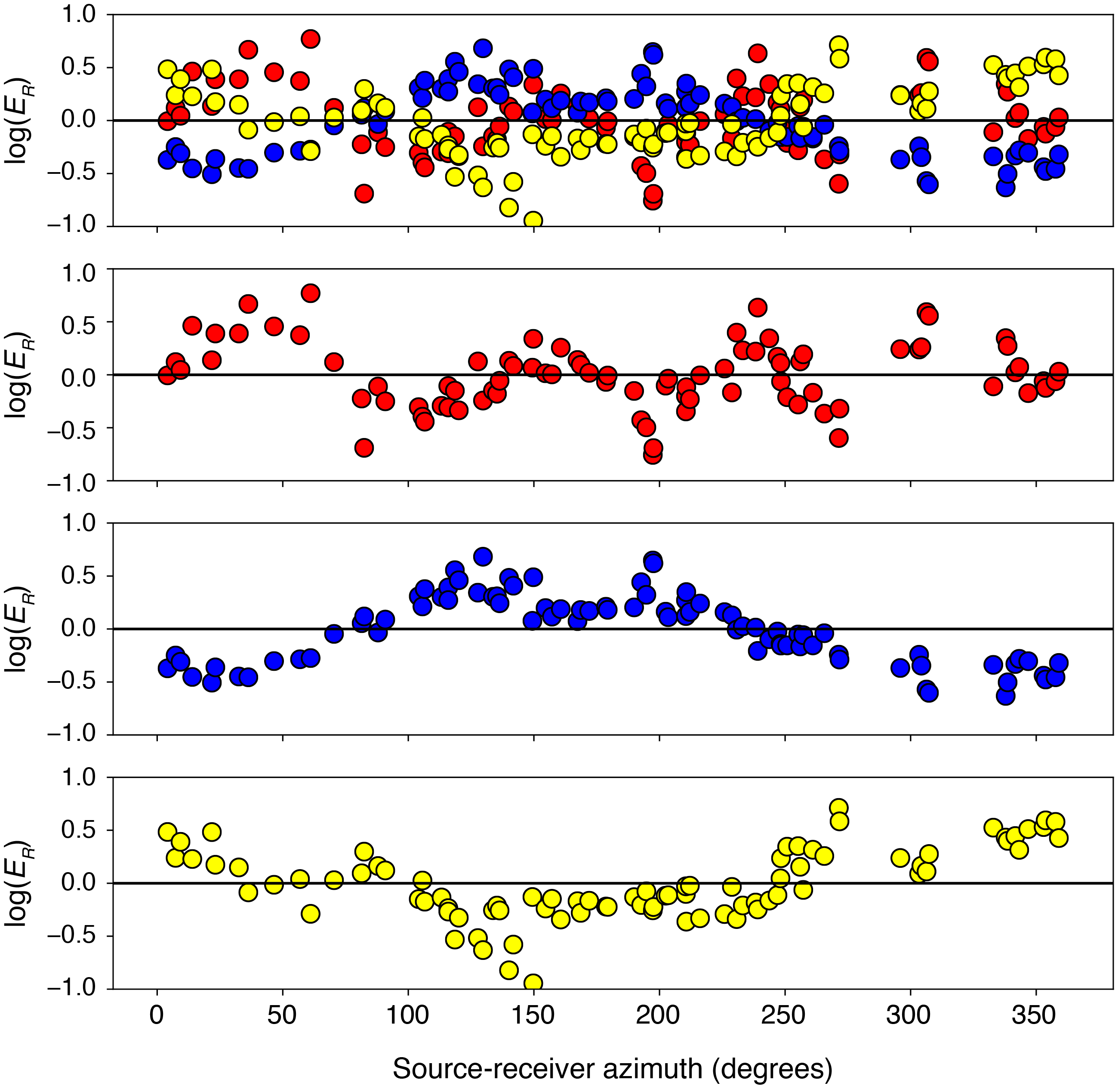}
\caption{Three-mode decomposition for the Cahuilla swarm. The modes correspond directly to the $w_k$ determined by fitting the GMM (Table \ref{table:cahuilla}).}
\label{fig:result_polar_cahuilla3}
\end{figure}

\begin{table}[]
\centering
\begin{tabular}{|l|l|l|}
\hline
Mode             & $w\,(K=2)$ & $w\,(K=3)$ \\ \hline
NW &  0.55 [0.49, 0.51] & 0.40 [0.30, 0.43] \\ \hline
SE &  0.45 [0.41, 0.60] & 0.34 [0.30, 0.45] \\ \hline
Bilateral     &         & 0.26 [0.21, 0.33] \\ \hline
\end{tabular}
\caption{Modal results for the Cahuilla swarm.}
\label{table:cahuilla}
\end{table}

The results of applying the method to the Cahuilla swarm data are shown in Figure \ref{fig:result_polar_cahuilla} and Table \ref{table:cahuilla}. In total, there are 829 earthquakes and 86 stations used for the final decomposition. First, we show the results for $K=2$. In this plot, each mode is assigned a different color. The elements of $\mu_k$ represent the average value of $\log{E_R}$ at a given station for mode $k$. Since there are two modes, each station has two different colored circles present. Value larger than zero indicate that $\log{E_R}$ at that station are amplified relative to the median (for that mode). $\mu_1$ and $\mu_2$ have peaks along the strike of the fault but in opposite directions. There is more than a factor of 15 difference in the $E_R$ between these opposing directions. Thus, the method has been able to identify the fault plane, and these modes represent unilateral directivity. For these two modes in the Cahuilla swarm (Table \ref{table:cahuilla}), $w_1=0.55$ and $w_2=0.45$. Since $w_1 \geq w_2$ in 95\% of the bootstrap samples, ruptures are statistically more likely to have NW directivity.

For the $K=3$ decomposition (Fig. \ref{fig:result_polar_cahuilla3}), two of the modes look very similar to the $K=2$ results. The third mode exhibits more complex behavior, with four peaks and four troughs. These peaks are essentially at conjugate orientations and suggest that there may be cross-faulting mixed in. Events have bilateral ruptures between 21-33\% of the time (Table \ref{table:cahuilla}). Thus it is the least likely of the three modes to occur.

\subsection{San Andreas (Creeping section)}

\begin{table}[htb]
\centering
\begin{tabular}{|l|l|l|l|}
\hline
Mode             & $w\,(K=2)$ & $w\,(K=3)$ \\ \hline
NW &  0.50 [0.46, 0.56] & 0.30 [0.26, 0.38] \\ \hline
SE &  0.50 [0.44, 0.54] & 0.24 [0.20, 0.40] \\ \hline
Bilateral     &         & 0.46 [0.26, 0.49] \\ \hline
\end{tabular}
\caption{Modal results for the Creeping section of San Andreas.}
\label{table:creeping}
\end{table}

\begin{figure}[htb]
\centering
\includegraphics[width=\columnwidth]{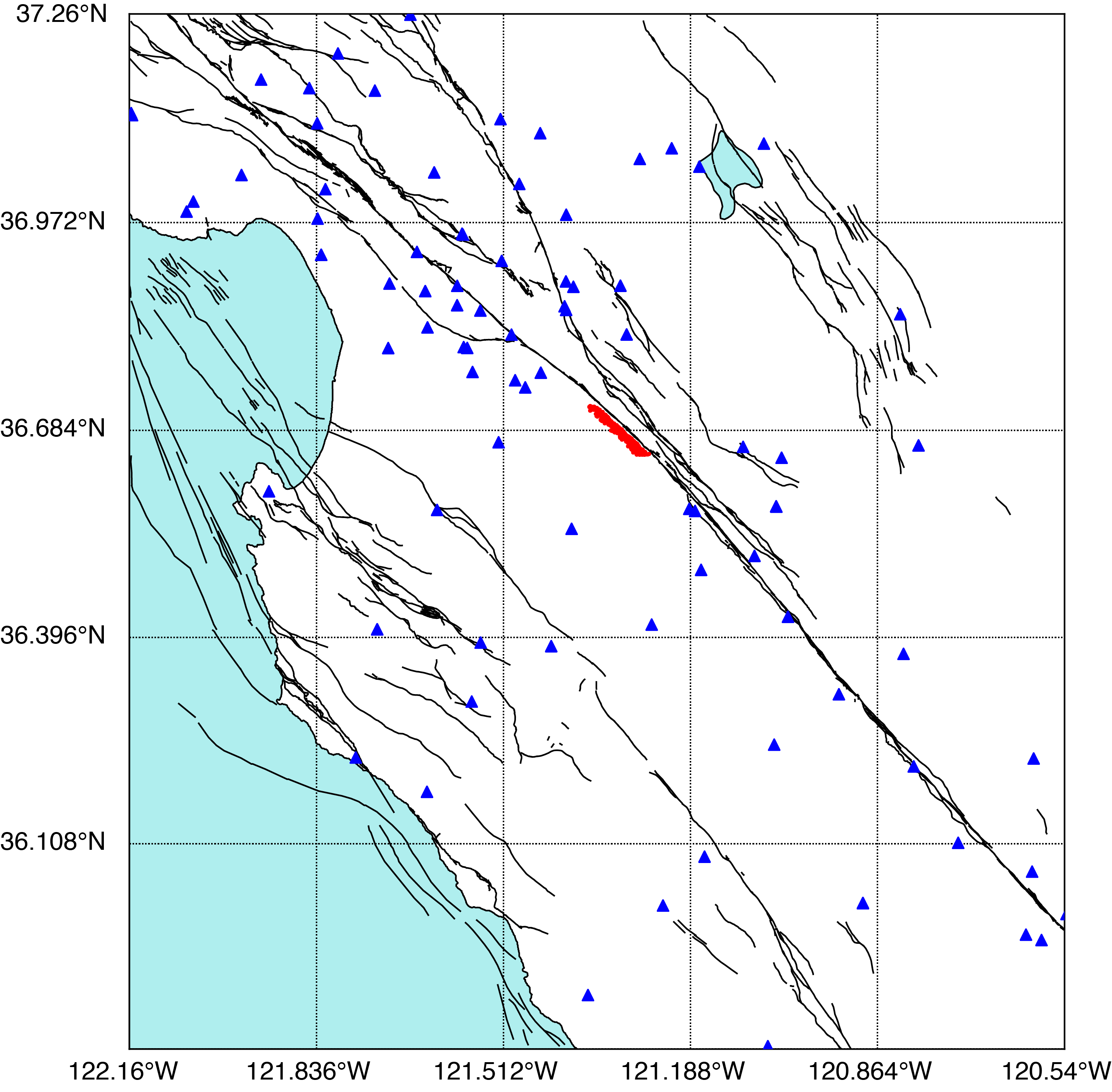}
\caption{Map of the creeping section of the San Andreas fault. Earthquakes are shown as red dots. Seismic stations used in this study are indicated by blue triangles.}
\label{fig:creeping_map}
\end{figure}

The second dataset analyzed in this paper is for the creeping section of the San Andreas fault. This area has a very active seismicity cluster that has produced 4118 events from 2002-2019 (Fig. \ref{fig:creeping_map}). All of the data are publicly available from the Northern California Earthquake Data Center. This cluster was studied by \citet{wang2011rupture} and serves as an additional baseline for our work.

\begin{figure}[htb]
\centering
\includegraphics[width=\columnwidth]{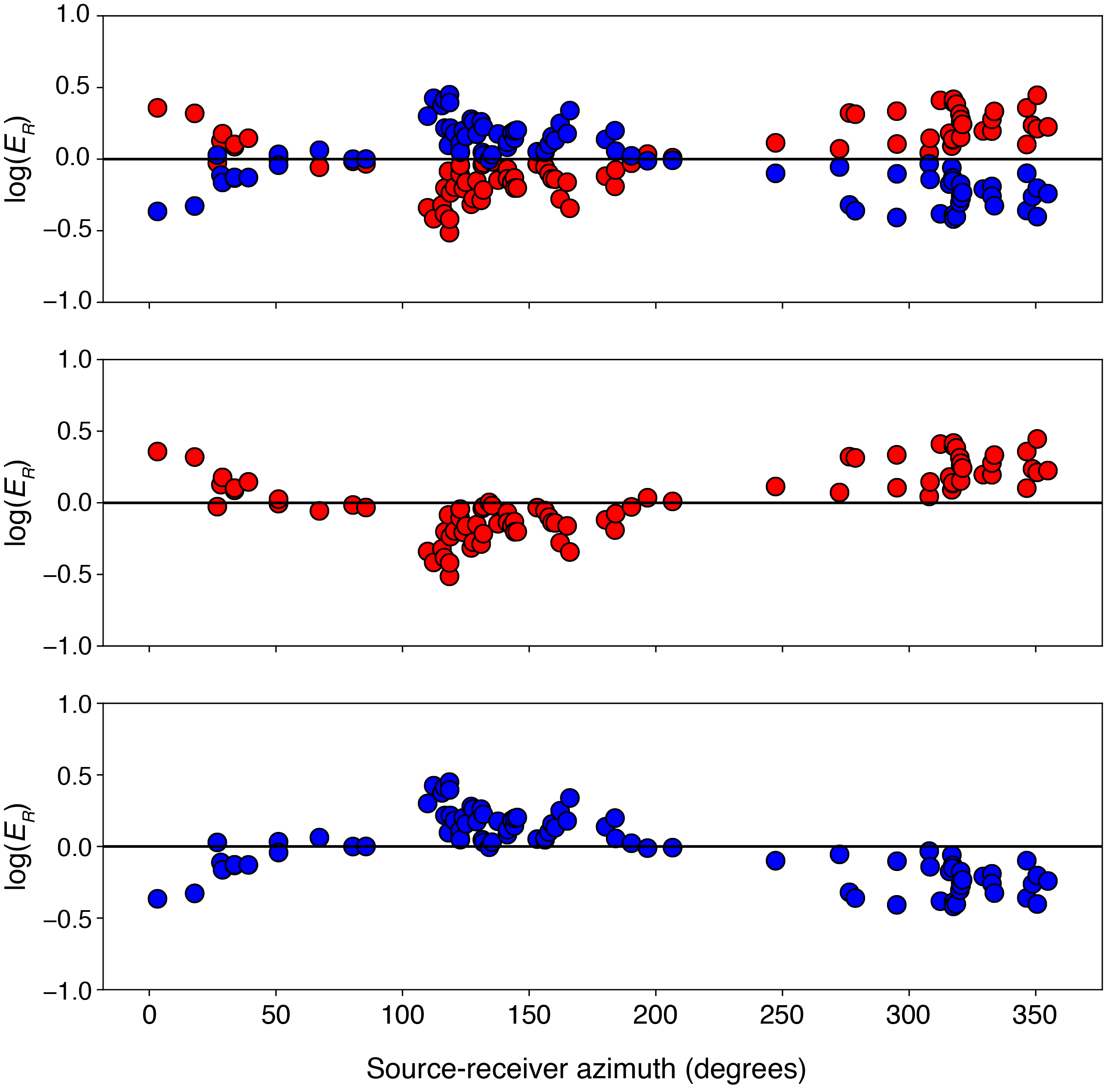}
\caption{Two-mode decomposition for the San Andreas fault (creeping section). The modes correspond directly to the $w_k$ determined by fitting the GMM (Table \ref{table:creeping}).}
\label{fig:result_polar_creeping}
\end{figure}

The $K=2$ results for the creeping section of the San Andreas fault are shown in Figure \ref{fig:result_polar_creeping} and Table \ref{table:creeping}. A total of 969 earthquakes and 90 stations were used to fit the model. As with the previous dataset, the fault plane is clearly identified for this segment, with prominent peaks in the NW azimuths for one mode and the SE azimuths for the other mode. The modes are evenly distributed on average and the confidence intervals further show that there is no evidence of a statistically preferred direction.

\begin{figure}[htb]
\centering
\includegraphics[width=\columnwidth]{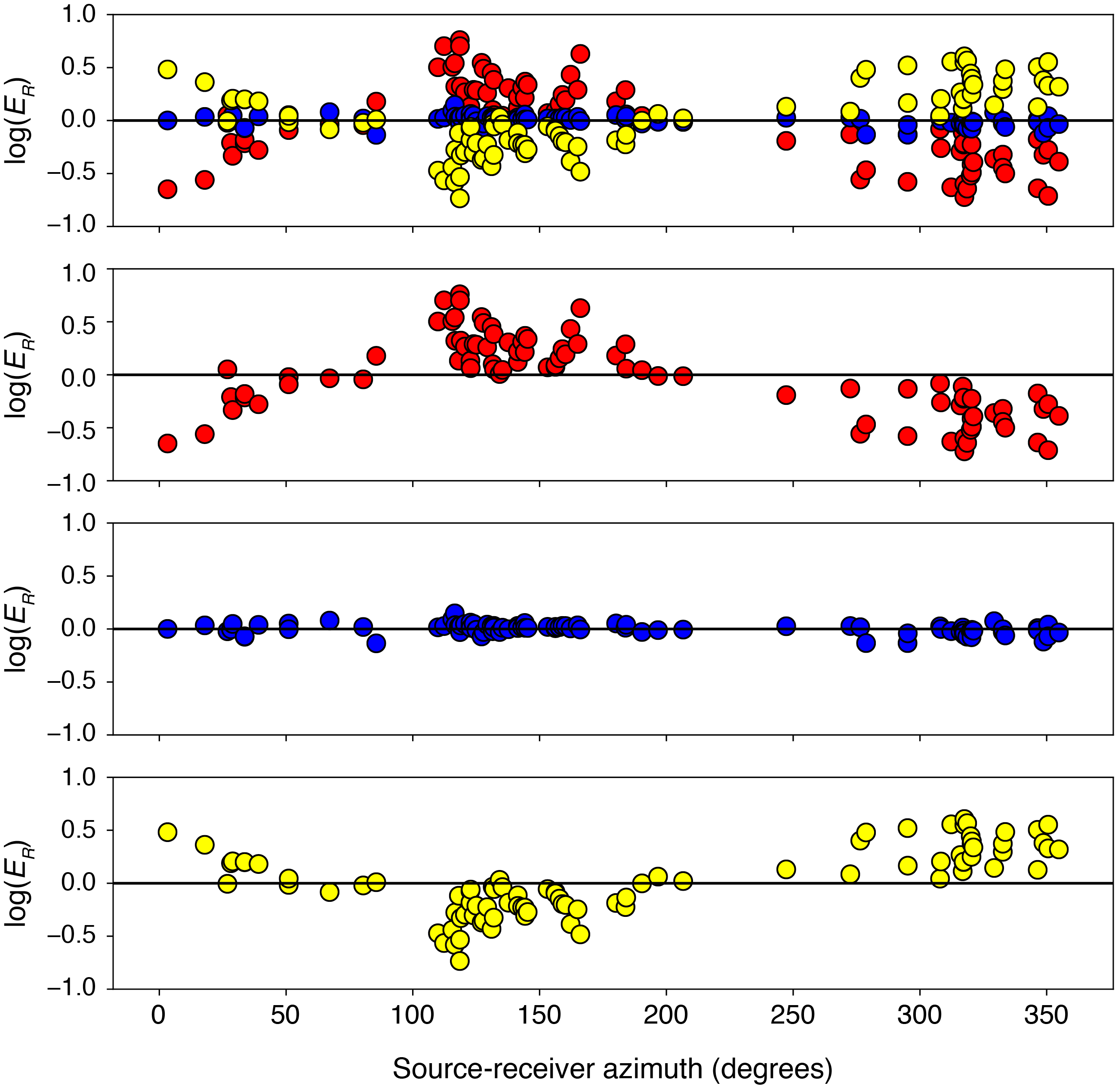}
\caption{Three-mode decomposition for the San Andreas fault (creeping section). The modes correspond directly to the $w_k$ determined by fitting the GMM (Table \ref{table:creeping}).}
\label{fig:result_polar_creeping3}
\end{figure}

For the $K=3$ decomposition, there are two unilateral modes (NW and SE) and one bilateral mode (Fig. \ref{fig:result_polar_creeping3}) the bilateral mode is the most frequent ($w=0.46$), and after factoring in the uncertainty, ruptures are unilateral 51-74\% of the time. \citet{wang2011rupture} found that about 40\% of the earthquakes on this segment had bilateral ruptures, which is similar to our observations. They also found that of the unilateral ruptures, SE ruptures were more common; however we find no evidence of this that is statistically significant.

\subsection{San Andreas (Parkfield)}
\begin{figure}[htb]
\centering
\includegraphics[width=\columnwidth]{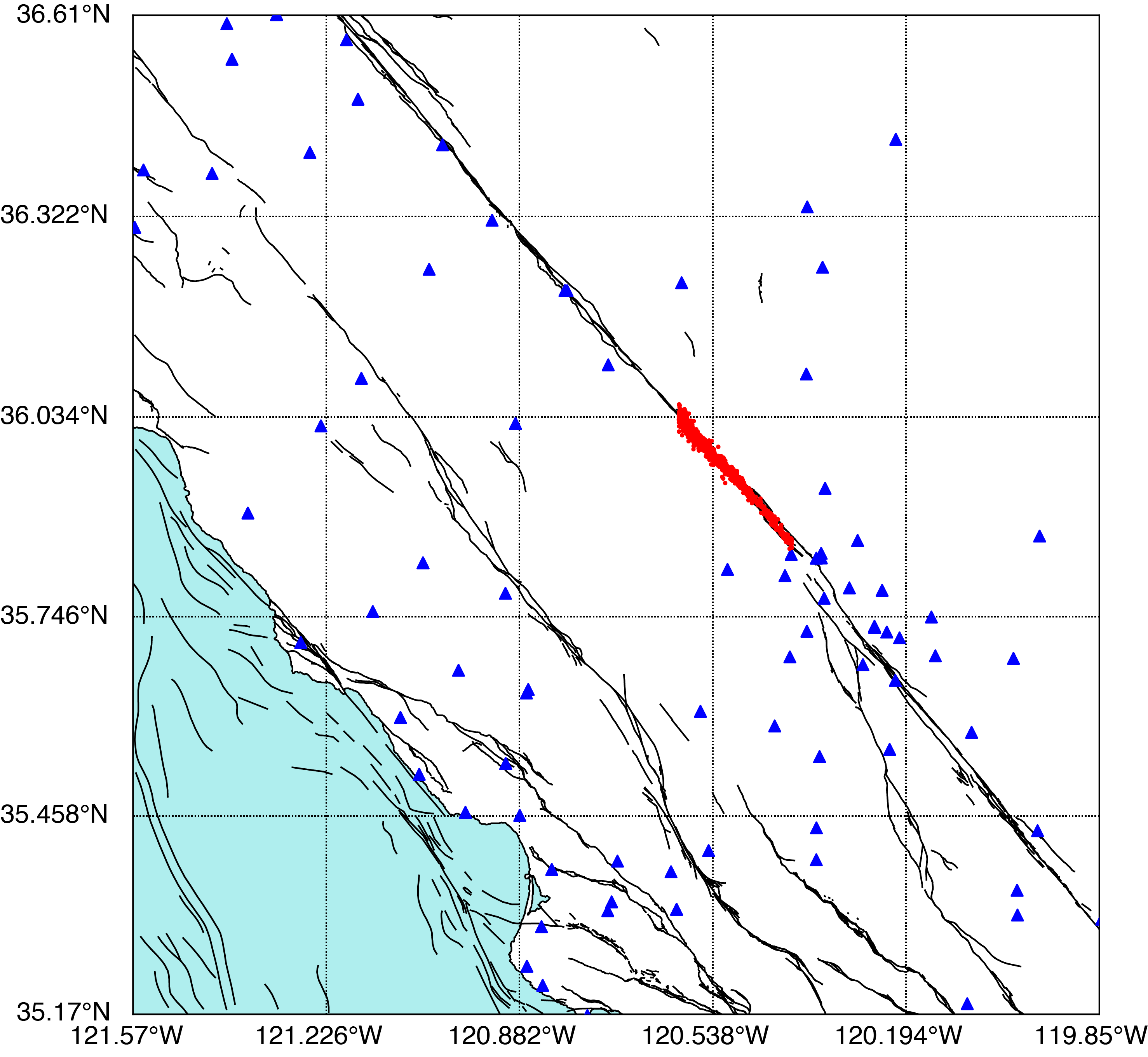}
\caption{Map of the Parkfield section of the San Andreas fault. Earthquakes are shown as red dots. Seismic stations used in this study are indicated by blue triangles.}
\label{fig:parkfield_map}
\end{figure}

The third dataset is for the Parkfield section of the San Andreas fault in central California (Fig. \ref{fig:parkfield_map}). We selected all 14,562 events in the NCEDC catalog that have occurred since 2002. This region was used for several directivity studies in the past \citep{kane2013rupture, lengline2011rupture} and serves as a point of comparison for our results. The earthquakes at Parkfield have very homogeneous focal mechanisms \citep{thurber2006three} which help to simplify the demonstrations of our approach.

\begin{table}[htb]
\centering
\begin{tabular}{|l|l|l|l|}
\hline
Mode             & $w\,(K=2)$ & $w\,(K=3)$ \\ \hline
NW &  0.63 [0.56, 0.68] & 0.21 [0.17, 0.42] \\ \hline
SE &  0.37 [0.32, 0.44] & 0.32 [0.28, 0.38] \\ \hline
Bilateral     &         & 0.47 [0.26, 0.51] \\ \hline
\end{tabular}
\caption{Modal results for the Parkfield cluster.}
\label{table:parkfield}
\end{table}

\begin{figure}[htb]
\centering
\includegraphics[width=\columnwidth]{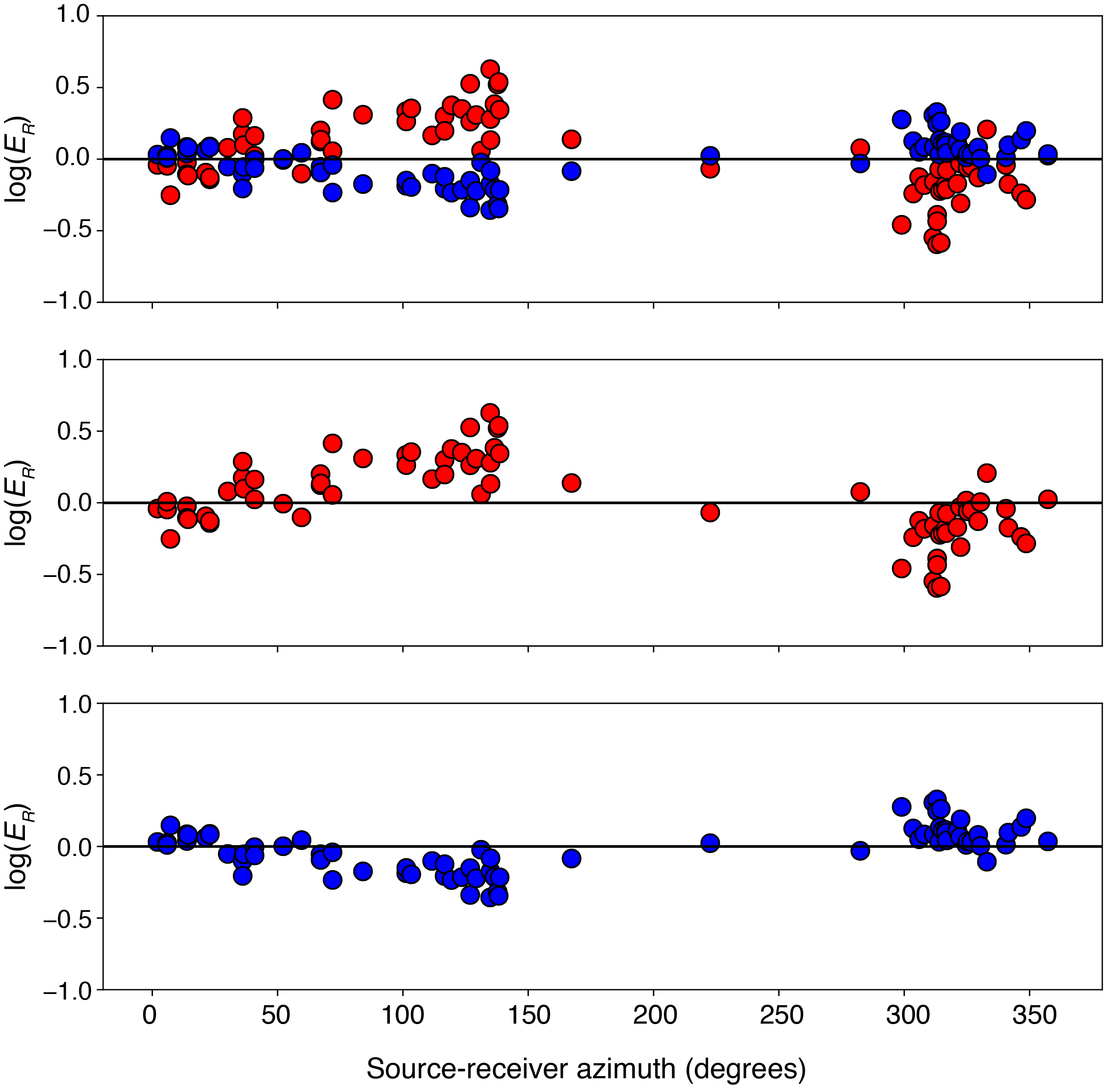}
\caption{Two-mode decomposition for the San Andreas fault (Parkfield section). The modes correspond directly to the $w_k$ determined by fitting the GMM (Table \ref{table:parkfield}).}
\label{fig:result_polar_parkfield}
\end{figure}

Our results for the Parkfield data $(K=2)$ are shown in Figure \ref{fig:result_polar_parkfield} and Table \ref{table:parkfield}. The model was fit to 618 events at 76 stations. The strike and fault plane are clearly identified, with prominent peaks in in the modes along the NW and SE azimuths. For the NW mode, $w=0.63$ with a one-tailed confidence interval of 0.56 at the 95\% level, which indicates that NW ruptures are statistically more frequent than SE ruptures. This finding contrasts the results of \citet{lengline2011rupture}, who concluded that most of the events had SE rupture directivity signals. 

\begin{figure}[htb]
\centering
\includegraphics[width=\columnwidth]{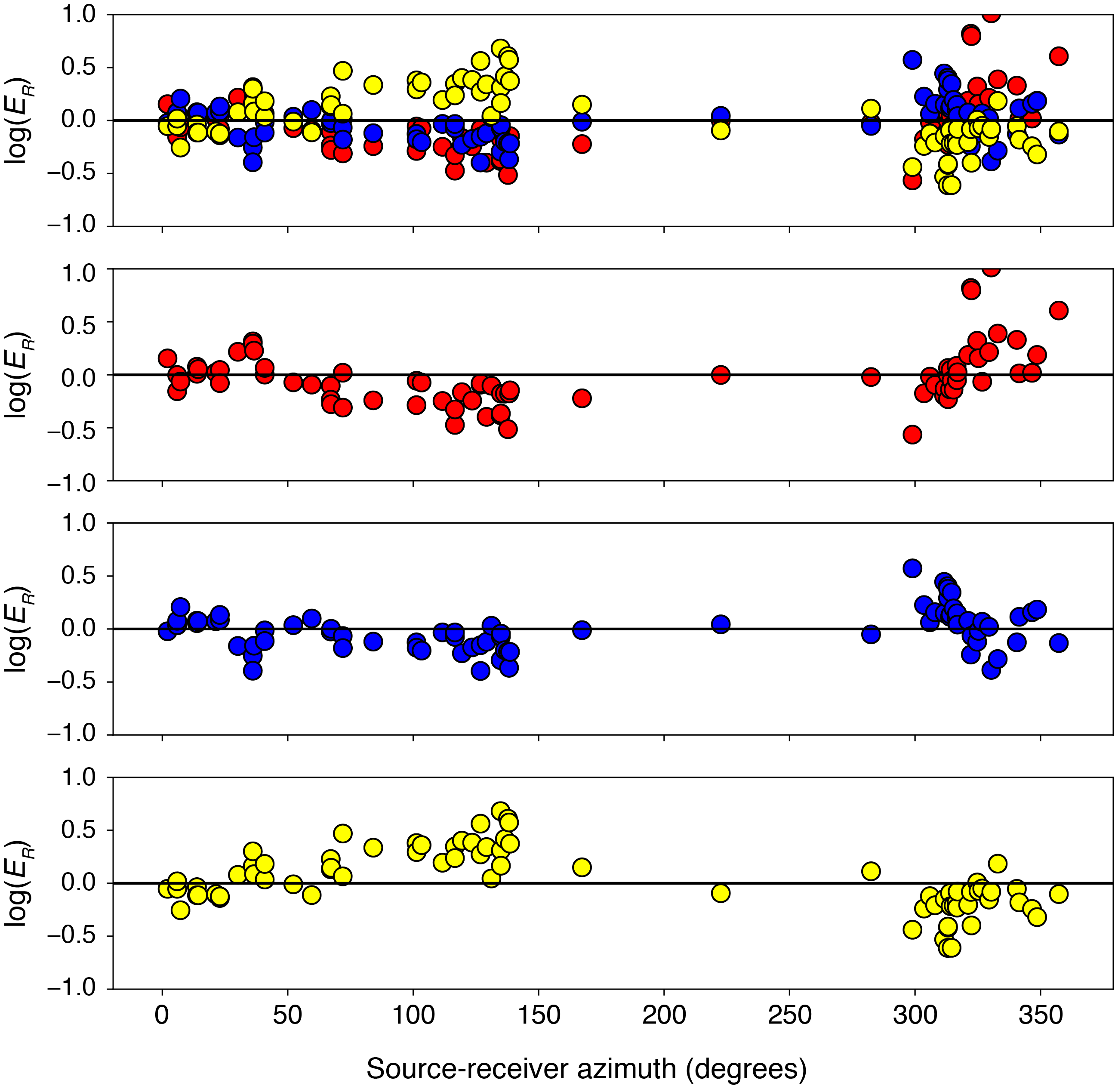}
\caption{Three-mode decomposition for the San Andreas fault (Parkfield section). The modes correspond directly to the $w_k$ determined by fitting the GMM (Table \ref{table:parkfield}).}
\label{fig:result_polar_parkfield3}
\end{figure}

In Figure \ref{fig:result_polar_parkfield3}, the results for the $K=3$ decomposition are shown. For this case, the NW and SE modes do not decompose as cleanly as in the $K=2$ case. This mixing makes it more difficult to confidently interpret the results, compared with the $K=2$ results. We find that the mode that appears the most bilateral occurs 26-51\% of the time.

\subsection{Hayward fault}

\begin{table}[htb]
\centering
\begin{tabular}{|l|l|l|l|}
\hline
Mode             & $w$ & 95\% CI \\ \hline
NW & 0.38 [0.33, 0.67] & 0.37 [0.16, 0.47]    \\ \hline
SE & 0.62 [0.33, 0.67] & 0.28 [0.17, 0.60]    \\ \hline
Bilateral     &        & 0.35 [0.06, 0.48]   \\ \hline
\end{tabular}
\caption{Modal results for the Hayward fault.}
\label{table:hayward}
\end{table}

\begin{figure}[htb]
\centering
\includegraphics[width=\columnwidth]{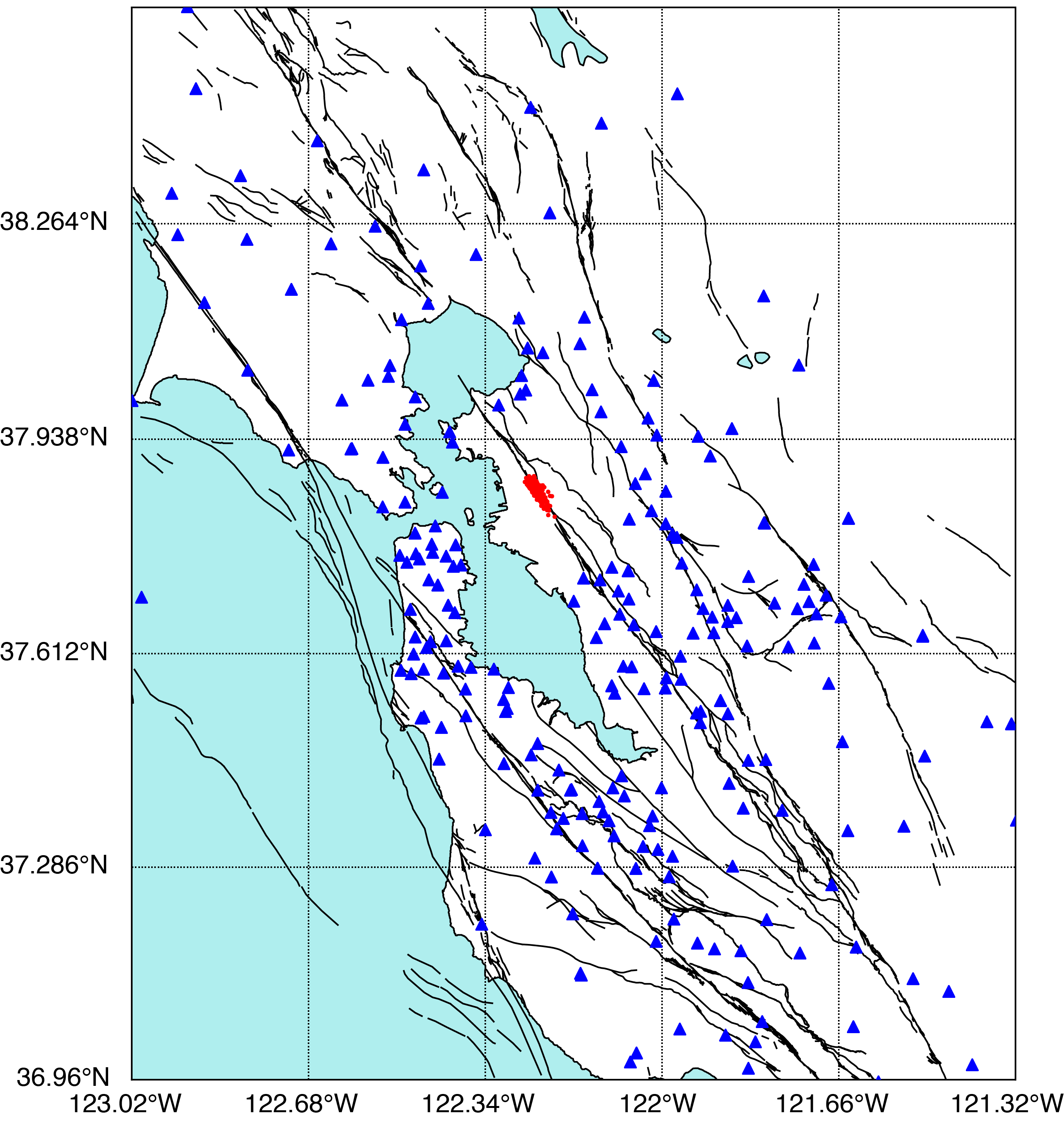}
\caption{Map of the Hayward fault and surrounding region. Earthquakes are shown as red dots. Seismic stations used in this study are indicated by blue triangles.}
\label{fig:hayward_map}
\end{figure}

The final dataset analyzed in this paper is for the Hayward fault. The seismicity here occurs over a long stretch of the fault, and so we chose a compact cluster to work with to ensure that the azimuths were similar to all events. There were 4118 events from 2002-2019 over the whole area and we used these to determine the source spectra (Fig. \ref{fig:hayward_map}). All of the data are publicly available from the Northern California Earthquake Data Center. The Hayward fault has a velocity contrast in the range 3-8\% \citep{allam2014seismic}.

\begin{figure}[htb]
\centering
\includegraphics[width=\columnwidth]{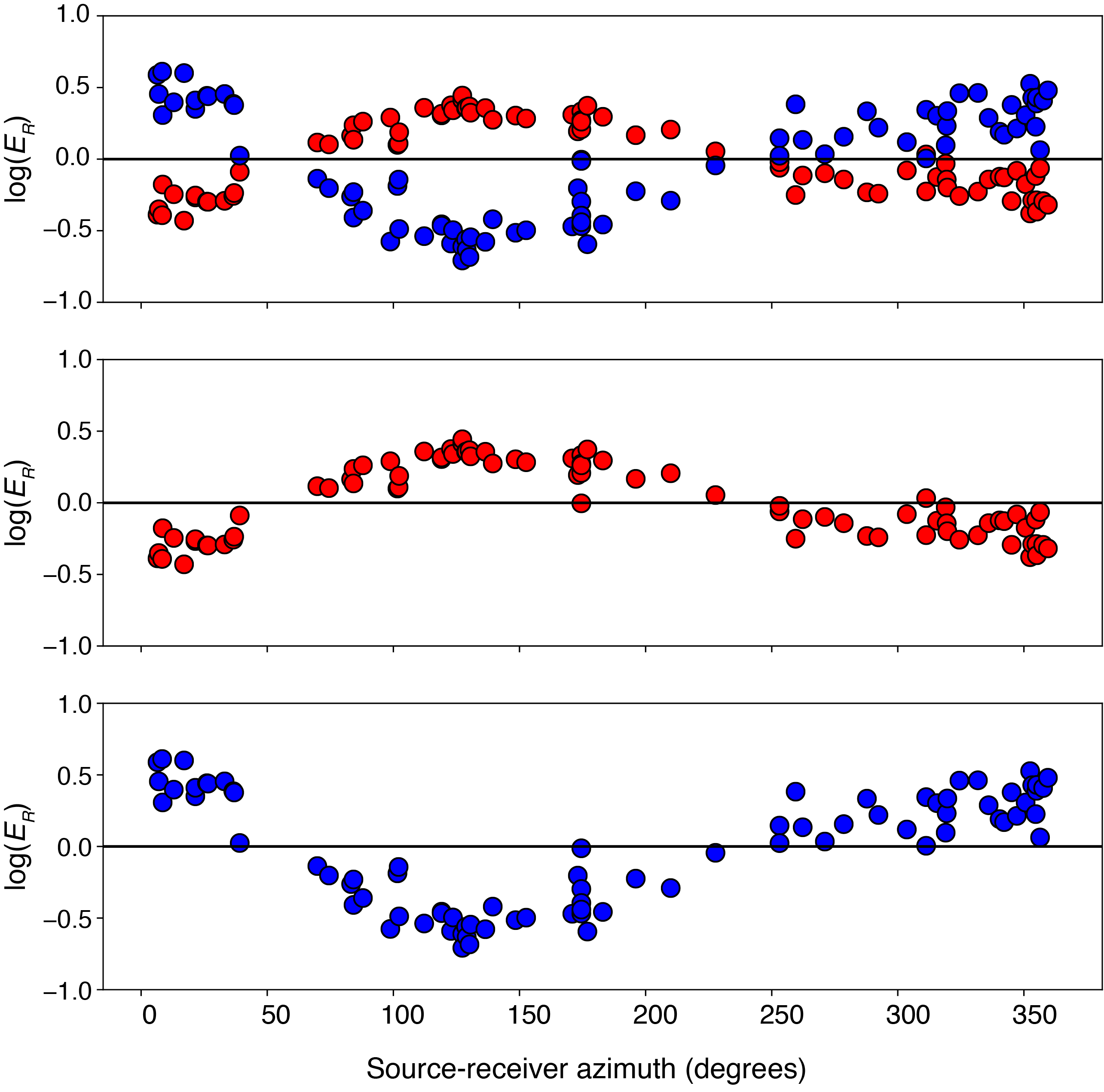}
\caption{Two-mode decomposition for the Hayward fault. The modes correspond directly to the $w_k$ determined by fitting the GMM (Table \ref{table:hayward}).}
\label{fig:result_polar_hayward}
\end{figure}

After all of the pre-processing, the Hayward dataset has 170 earthquakes at 82 stations. We show the results for $K=2$ in Figure \ref{fig:result_polar_hayward} and Table \ref{table:hayward}. As with the previous study areas, here the data separate clearly into two modes with directivity behavior in NW and SE directions, which aligns with the strike of the fault. The $w$ values for the modes vary significantly for the best-fitting values, but interestingly, have exactly the same confidence intervals. Thus, they are statistically indistinguishable, albeit with large uncertainties.

\begin{figure}[htb]
\centering
\includegraphics[width=\columnwidth]{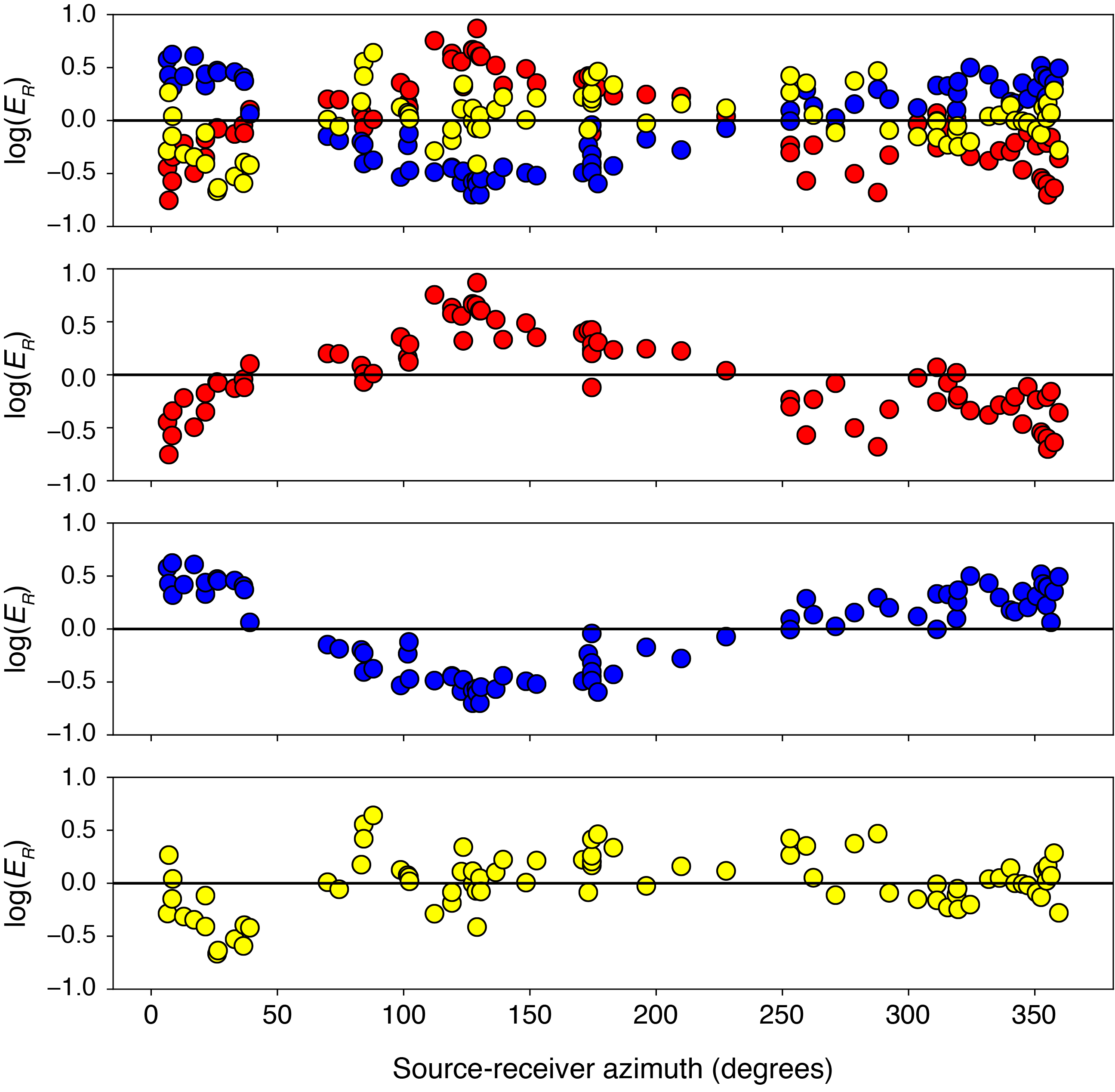}
\caption{Three-mode decomposition for the Hayward fault. The modes correspond directly to the $w_k$ determined by fitting the GMM (Table \ref{table:hayward}).}
\label{fig:result_polar_hayward3}
\end{figure}

When applying a $K=3$ model to the data (Fig. \ref{fig:result_polar_hayward3}, the modes are distinct and form smoothly varying patterns as a function of azimuth. The three modes also have sizable uncertainties as estimated by the bootstrapping, but we can say that unilateral modes occur between 52-94\% of the time; thus they are the most frequent mode of rupture.

\section{Discussion}

\subsection{The importance of studying directivity modes}
Directivity modes characterize the first-order kinematics of rupture propagation during earthquakes. It is important to understand these modes and their statistical tendencies because they can provide valuable observational constraints on the physics of earthquakes. One of the important findings of this study is that for the four examined fault segments, unilateral ruptures are more frequent than bilateral ruptures. Specifically, the best-fitting rates vary from 53-74\% over the four regions tested, and the uncertainty estimates show that this principal conclusion is robust. We find evidence of a statistically preferred rupture direction in two regions: the Cahuilla swarm and Parkfield. For the other regions, the uncertainties on the $w_i$ values are larger than the differences between them. This does not mean a preferred direction does not exist; rather the question could potentially be addressed better in the future by incorporating more events, which may tighten the confidence intervals. In all cases, however, there is no evidence of a single unilateral rupture direction being overwhelmingly likely.

Another noteworthy finding is that for these fault segments, there is a clear preference for ruptures to propagate horizontally more often than vertically. As discussed below, along-dip ruptures will generally be included in the bilateral mode because there is little sensitivity to them. This means that our conclusions about the predominance of unilateral ruptures not only applies with respect to bilateral cases, but also along-dip cases. Such a finding was made by \citet{mcguire2002predominance}, who observed that most of these were unilateral ruptures. For earthquakes large enough to rupture the full seismogenic zone, along-strike ruptures are expected purely from a geometrical perspective, although this says nothing about whether they should be unilateral or bilateral. However for small earthquakes, our observations may be unexpected. They further suggest that there are attributes of the faults that are breaking the symmetry and leading to horizontal ruptures being so prevalent. 

An additional reason that these findings are important is because observational studies of earthquake source properties commonly assume that rupture areas are circular \citep[e.g.][]{calderoni2012stress, abercrombie2015investigating, ross2016toward}. However, the geometry of the rupture area has an important influence on quantities like stress drop and radiated energy \citep{kaneko2015variability}. Our results suggest that for the examined datasets, unilateral ruptures are more frequent than bilateral ruptures, and this may indicate that some of the large scatter commonly observed in source properties may result from assuming the wrong rupture geometry.

One physical explanation for the systematic tendencies for unilateral directivity in earthquakes is from rate- and state-dependent friction. When the nucleation size $h$ is much smaller than the potential seismogenic region, this results in ruptures that tend to propagate unilaterally \citep[e.g.][]{michel2017pulse,lin2018microseismicity}. However the expected rates of unilateral ruptures for this type of physical model have not been analyzed rigorously. Future work in this area could be helpful to compare with observations.

Directivity modes are also important from a hazard perspective because the rupture propagation pattern has a significant effect on the strong ground motion. While earthquake stress drop is widely accepted to influence ground motion amplitudes in certain contexts \citep[e.g.][]{trugman2018strong,oth_relation_2017,baltay_uncertainty_2017}, directivity has received less attention to date but is readily acknowledged to be one of the few additional source parameters that does have a major impact \citep{bozorgnia_nga-west2_2014,douglas_recent_2016}. Therefore being able to provide some expectations on the possible range of modal probabilities can better inform hazard estimates.

\subsection{Data-driven approaches to studying directivity}

There are a variety of advantages to studying directivity with a data-driven approach, as opposed to using techniques that fit kinematic rupture models. The first is that a single model can be fit to the data for all earthquakes simultaneously. This allows for weaker signals to be extracted from the data because the information that is common to all events can be averaged, which acts to suppress noise. Traditional approaches generally fit a model separately to each event, and then if performed for enough events, then the results can be averaged for some statistical estimate. However this type of approach is limited by the capabilities of fitting a model to individual events.

Another reason is that there are generally fewer assumptions necessary with a data-driven approach, compared with traditional model fitting. For example, we do not need to assume that our physical model is correct, or that we understand the statistical properties of the noise. However, as in most analyses of earthquake directivity, we assume that azimuthal variations in seismic spectra are caused primarily by source effects, rather than 3D variations in path effects. While this is likely a valid approximation to first order, in reality the observed spectra will contain hints of both effects.

\subsection{Assumptions and interpretability of results}

In this study, there are several key assumptions that underlie the analysis. The first, and arguably the most important, is that we assumed the number of distinct modes of rupture propagation a priori. By making an assumption that a certain number of modes exist in the data, we are able to search for and identify their centroids $\mu_k$ and variances. If these assumptions are violated, for example in the case that there are more than three distinct modes, then the resulting $\mu_k$ vectors will be complicated and uninterpretable. In all four datasets that we tested here, the $\mu_k$ are generally simple, smoothly varying functions of azimuth, which suggests that the assumptions are justified. However when applying the data to other datasets, such as clusters with more than one dominant fault strike, more care will be needed to determine the optimal number of modes in the data. There are various strategies for determining this objectively, including the use of information criteria or silhouette analysis. We applied a silhouette analysis to the data and in all regions, the metric favored $K=2$. However, as we have shown, $K=3$ models provide additional insights into the directivity physics and we believe that examining both scenarios together provides a more informed result.

If the recovered modes reflect approximately end-member rupture scenarios, one way of assessing the degree of unilateral directivity for a given event \citep[e.g.][]{boatwright2007persistence} is by calculating the probability that an event belongs to the best-matching mode. This could be useful for identifying the events with the strongest directivity signals in an objective manner. However, it should be mentioned that this strategy will only determine whether an event best matches the centroid of the mode; if an event has even stronger directivity signals than the centroid, the probability will be diminished in proportion to the deviation from the centroid. This simply means that events with the highest probability values will not be the events with the strongest directivity.

Another important aspect of the results is that the modal shapes which we have identified as bilateral are a bit more complicated in practice. In particular, if a rupture is purely along-dip, then without stations right on top of the rupture, it is very difficult to observe any directivity patterns. These ruptures will appear as generally symmetric directivity patterns and will likely be aggregated into the bilateral mode. Thus, the bilateral mode (particularly the $w_k$ value associated with it) should be interpreted as an upper bound for the likelihood of bilateral ruptures. It should be understood that this results from a lack of sensitivity to along-dip ruptures. Some studies have included other information, such as depth phases \citep[e.g.][]{he2017rapid}, to help constrain these cases, but in general it is a challenge common to all directivity analyses.

The results in the paper do not depend on the particular model used (e.g. GMM). We tested several other latent variable models including BIRCH and spectral clustering, and when setting the number of clusters to $K=2,3$, recover modes that look very similar.

\section{Conclusions}
We developed a new approach to resolving modes of directivity in large earthquake populations. We formulate the problem as one of recovering $K$ latent variable modes from the azimuthal energy distributions of many earthquakes, where each mode is a distinct state of rupture propagation. A gaussian mixture model is used to determine these modes, which allows for simultaneous estimation of the fraction of events that best align with each mode of rupture propagation. In the process, we have not needed to fit a kinematic directivity model to the data; the decomposition is possible because the data exhibit this type of structure naturally. We applied the method to four large earthquake clusters, and performed an assessment of the uncertainties for each mode of rupture, for each dataset. Our results indicate that unilateral ruptures are more likely to occur than bilateral ruptures for the examined datasets, even after incorporating the uncertainties.

\section*{Acknowledgments}
A. Anandkumar is supported in part by Bren endowed chair, Darpa PAI, Raytheon, and Microsoft, Google and Adobe faculty fellowships
\bibliography{main}
\bibliographystyle{icml2019}

\end{document}